# Efficient Training of the Memristive Deep Belief Net Immune to Non-Idealities of the Synaptic Devices

*Wei Wang, Barak Hoffer, Tzofnat Greenberg-Toledo, Yang Li, Minhui Zou, Eric Herbelin, Ronny Ronen, Xiaoxin Xu, Yulin Zhao, Jianguo Yang, and Shahar Kvatinsky\**

W. Wang, B. Hoffer, T. Greenberg-Toledo, Y. Li, M. Zou, E. Herbelin, R. Ronen, and S. Kvatinsky
The Andrew and Erna Viterbi Faculty of Electrical and Computer Engineering, Technion-Israel Institute of Technology, Haifa, Israel 3200003
Email: wei.wang@campus.technion.ac.il
Email: shahar@ee.technion.ac.il
X. Xu, Y. Zhao, J. Yang
Institute of Microelectronics, Chinese Academy of Sciences, Beijing, 100029, China

The tunability of conductance states of various emerging non-volatile memristive devices emulates the plasticity of biological synapses, making it promising in the hardware realization of large-scale neuromorphic systems. The inference of the neural network can be greatly accelerated by the vector-matrix multiplication (VMM) performed within a crossbar array of memristive devices in one step. Nevertheless, the implementation of the VMM needs complex peripheral circuits and the complexity further increases since non-idealities of memristive devices prevent precise conductance tuning (especially for the online training) and largely degrade the performance of the deep neural networks (DNNs). Here, we present an efficient online training method of the memristive deep belief net (DBN). The proposed memristive DBN uses stochastically binarized activations, reducing the complexity of peripheral circuits, and uses the contrastive divergence (CD) based gradient descent learning algorithm. The analog VMM and digital CD are performed separately in a mixed-signal hardware arrangement, making the memristive DBN high immune to non-idealities of synaptic devices. The number of write operations on memristive devices is reduced by two orders of magnitude. The recognition accuracy of 95%~97% can be achieved for the MNIST dataset using pulsed synaptic behaviors of various memristive synaptic devices.

# 1. Introduction

The separation of memory and computing units in the conventional von-Neumann architecture computing systems, which causes the memory wall bottleneck, is the main issue preventing the artificial neural network from competing with the human brain in efficiency and in intelligence.[1,2] Emerging non-volatile memory devices which have tunable resistance, i.e., memristive devices, including resistive random-access memory (RRAM),[3,4] phase-change memory (PCM),[5] ferroelectric random-access memory (FeRAM), etc. are promising techniques to solve the memory wall issue.[6–8] They can store information in an analog way and process information at the same location, acting as artificial synaptic devices and enabling in-memory computation just like what happens in the human brain.[9,10] Furthermore, an array of memristive devices can efficiently perform the vector-matrix multiplication (VMM), which is in the computational kernel of a deep neural network (DNN), via Ohm's law and Kirchhoff's current law in one step,[11–13] making it a promising way to greatly accelerate the DNN and to power future artificial intelligence.[14–16]

However, there are several remaining issues before the promise comes true. Firstly, since the memristive VMM operations are conducted in the analog domain, expensive analog-to-digital and digital-to-analog converters (ADCs and DACs) and additional circuits for the neuron's non-linear activation functions are needed for the communication in adjacent layers of the DNNs.[17,18] To avoid the use of high precision and expensive ADCs and DACs, novel spiking rate-coded neurons have been proposed.[19,20] However, the spiking rate-coded neuron circuit is still complex and informational inefficient.[21] Secondly, the online training is usually performed by tunning the conductance of the synaptic devices in a closed-loop write method (iteratively write and verify to achieve the target value for a single weight update request[14,22]), which is inefficient. The open-loop method which can potentiate the synaptic weight by a single write pulse and depress the weight by another single write pulse in the opposite direction is preferred.[23,24] However, this method fades due to that the neural network performance is greatly degraded by the non-idealities of the memristive synaptic devices.[25,26] Particularly, process variation and stuck-at fault errors have been widely reported to cause performance degradation, although can be partially compensated by various methods with extra costs[27–33][34–39]. These issues can be generally attributed to the non-biological conventional learning algorithm of DNN, that is, error backpropagation based gradient descent weight update,[40,41] which needs both the VMMs and the conductance tuning in high precision.[42–44] Novel neural network structures and learning algorithms need to be explored to address these issues.

In this paper, we investigated the hardware implementation of the memristive deep belief net (DBN) based on the learning algorithm of contrastive divergence (CD).[45] The memristive DBN is composed of stacked restricted Boltzmann machines (RBMs),[46] where the VMM operations have binary inputs and stochastically binarized outputs, needing no ADCs or DACs in the peripheral circuits. The RBM is trained by accumulating the CD in a separated digital array and updating the synaptic weights periodically via the open-loop write method on the memristor array. The training of the DBN needs no additional cache memory to store the intermediate states of hidden layers, nor dedicated circuits for non-linear activation functions. The proposed memristive DBN shows high immunity to non-idealities of the synaptic devices, greatly relaxing the specifications for memristive synaptic devices in multiple dimensions.

## 2. Network structure and hardware design
### 2.1. The DBN and RBMs

The structure of the investigated DBN is shown in **Figure 1a** which consists of three stacked RBMs[45]. Each RBM has a visible layer, a hidden layer, and a weight matrix ($w$) connecting them. For supervised learning tasks, taking MNIST dataset as an example in this work,[40] the images are fed into the visible layer of the first RBM, and the labels are part of the visible layer of the last RBM. Unlike conventional DNNs based on error backpropagation algorithms, the DBN rely on the consecutive training of each RBM via the CD algorithm.

The CD is obtained by alternative Gibbs sampling between the visible layer and the hidden layer within each RBM which requires both forward and backward VMMs as well as binary sampling operations. All input signals are binary digital signal which can be easily generated by digital circuits. For instance, in the first RBM layer (RBM 1, Figure 1a and 1b), each image in the MNIST dataset was binarized (pixel value to be either '0' or '1') and converted to a one-dimensional vector as the states of the visible neurons ($v$). After alternative Gibbs sampling (see Methods for more details), the hidden neuron states ($h$), the reconstructed visible neuron states ($v'$), and the reconstructed hidden unit states ($h'$), which are the local information needed to calculate the CD and update the weight matrix, are obtained.

After the first RBM layer is trained, the state of the hidden neurons ($h$) will be the input of the second RBM layer (RBM 2 in Figure 1a) for its training. The last RBM layer (RBM 3) takes both the states of the hidden neurons of the previous RBM layer and the label vector ($l$) as the input (see Methods for more details). The weight matrix in the RBM 3 is partitioned into two parts ($w_3$ and $w_4$ for clarification). The learning of DBN by consecutive training of the stacked RBM layer is named as "greedy learning" method.[45]

In a conventional DNN with the error backpropagation algorithm, the error propagated from the last layer would gradually vanish which makes it harder for hardware implementation. Additionally, the gradient descent of the weight relies on both the input of the layer and the error back propagated from the last, which raises the issue of the data dependency. In other words, the states of the neurons in all layers need to be stored before the backpropagated error arrives and the weight is updated. Whereas, in the DBN, all the neuron states are binarized ('0' or '1') and the CD elements are ternary values ('-1', '0', and '1', see Eq. 10 in Methods) making them easier to be processed by the hardware. Moreover, the calculation of the gradient descent, i.e., the CD, depends only on the local information of the neurons, further simplifying the memory requirements and hardware design in the training stage.

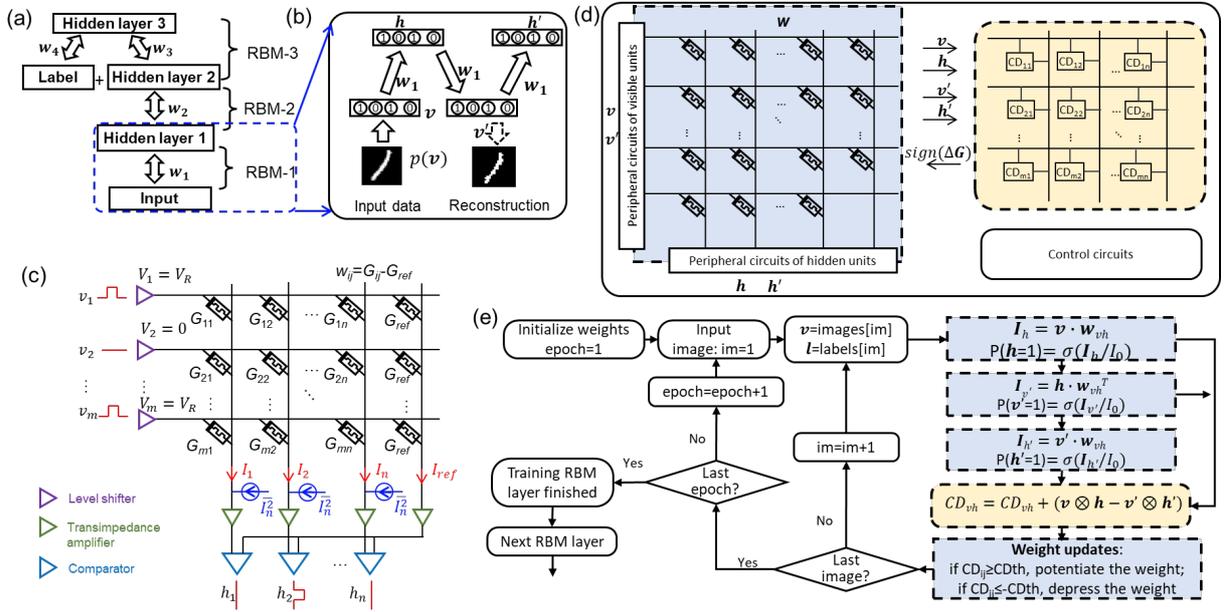

**Figure 1. The structure of the memristive DBN and the mixed-signal design of memristive RBM.** a) The structure of a typical DBN for the training of MNIST dataset consisting of three restricted Boltzmann machines (RBMs). b) Illustration of Gibbs sampling between the visible layer and hidden layer in a single RBM during the training. c) Forward vector-matrix multiplication (VMM) in a memristive crossbar array and binary sampling in the outputs to implement the Gibbs sampling from the visible neurons to the hidden neurons. d) Design of mixed-signal training of single RBM layer in the deep belief nets. e) Flow chart of the training of a single RBM layer in the greedy learning algorithm of the DBN. The light-blue and light-yellow colored blocks are the procedures of analog operations (VMM, stochastic sampling, and weight updates) and digital operations (CD calculation and accumulation), respectively, which are conducted by the components in (d) with the same colors.

## 2.2 Implement VMM with in-situ stochastic activations

The Gibbs sampling operation in an RBM can be fully hardware-performed by the memristive crossbar array with an additional noise current in each output node as shown in Figure 1c. Figure 1c performs the forward VMM and output sampling from the visible neurons to hidden neurons (Eq. 6 and Eq. 8). The binary states of the visible neurons (input digital signal) are converted to the read voltage ($V_i$, $i \in 1, 2, ..., m$) as the input of the memristive array with the size of m-by-n, which performs the VMM via Ohm's law and Kirchhoff's current law. Since the input of the VMM operation is a binarized vector, only a level shifter is needed (i.e., 1-bit DAC). The current output of the memristive array can be denoted as

$$I_j = \sum V_i G_{ij}, \tag{1}$$

where $j \in 1, 2, ..., n$ is the column index of the memristive array, and $G_{ij}$ is the conductance of the device in the *i*th row and *j*th column. A separate column of the memristive device with fixed reference conductance ($G_{ref}$) is used to provide the reference current,

$$I_{ref} = \sum V_i G_{ref}. \tag{2}$$

A noise current $[I_{noise} \in \mathcal{N}(0, I_n^2)]$ is injected into each output node of the memristive array. The output current is then converted to the voltage by a trans-impedance amplifier (TIA) and compared with the voltage output of the reference column by a comparator (i.e., 1-bit ADC). The hidden neuron states thus can be written as

$$h_j = \begin{cases} 1, & I_j - I_{ref} \geq I_{noise} \\ 0, & I_j - I_{ref} < I_{noise} \end{cases}, \tag{3}$$

which reproduces Eq. 7 with the weight $w_{ij}$ being $G_{ij}$-$G_{ref}$. Similarly, the backward VMM and output sampling from the hidden neurons to visible neurons (Eq. 8) can be implemented by placing the read circuits in the hidden neurons and the noise currents, TIAs, and comparators in the visible neurons. Note that the current design only supports sigmoid-type activation function which is the only activation function needed in the memristive RBM and DBN. Supporting of other activation functions that are needed in other neural networks requires further investigation. We have separately simulated and verified the circuit functionality of the noise current generation, trans-impedance amplifier, comparator, level shifter. However, simulation of the CMOS peripheral circuit for a specific technology, including specific limitations, such as operational voltage, parasitic capacitance, to explore the bandwidth and latency of the design memristive RBM and DBN, needs further investigation and should be the next step of the work.

The stochasticity of visible or hidden neurons can also be provided by the intrinsic read noise of the memristive device by properly tuning the signal-to-noise ratio,[47] which can further simplify the hardware implementation of the DBN. Here, we utilize the external noise such that we can turn the noise current off for fast inference.

**2.3 Memristive array and CD accumulation array**

To enable the learning of memristive DBN tolerant to non-idealities of synaptic devices, we used a mixed-signal hardware design of the RBM layer composed of an analog memristor array and a signed digital counter array (Figure 1d). The memristor array is composed of a crossbar array of memristors with the conductance of $G_{ij}$ and reference cells with the conductance of $G_{ref}$. Only the memristor array participates in the VMM as detailed in Figure 1c. The forward and backward VMMs and stochastic excitation result in two sets of binarized visible neuron states and hidden neuron states (*v* and *h*, *v'* and *h'*). The digital counter array will perform the outer product calculation of the CD matrix ($CD_{ij}=v_ih_j-v'_ih'_j$) and accumulate the ternary CD values in its cell, i.e., a signed digital counter. An identical pulse will be applied to the memristor cell to potentiate or depress its weight ($G_{ij}$) when the corresponding $CD_{ij}$ in the digital array reaches a threshold ($\geq CD_{th}$) or below the negative threshold ($\leq -CD_{th}$). This will result in a positive or negative conductance change ($\Delta G$) on the memristor cell defined by the memristive synaptic behavior. No verifying read operations will be needed. The training procedure of the memristive RBM is shown in Figure 1e, where the analog VMM and neuron state sampling steps, as well as the weight updates, are light-blue colored and the digital CD calculation and accumulation step is light-yellow colored, corresponding to the colored components in Figure 1d.

The proposed mixed-signal approach for memristive DBN training shares some similarities with the state-of-the-art techniques recently proposed to improve the training performance of the deep neural network,[48,49] however, also shows distinct features. S. Ambrogio et al.[48] proposed a hybrid synaptic cell composed of non-volatile memristive devices and volatile capacitor gated transistors (2PCM + 3T1C) for a DNN implementation. The capacitor gated transistor branch of the synaptic cell has high linearity for weight updating and performs both VMMs and weights updates. The accumulated weights updates were transferred to the non-volatile memristive devices periodically. Here, in our proposed neural network, the CD counter array only accumulates gradient descent (weight update request), and the memristive array performs the VMMs alone. S. R. Nandakumar et al.[49] proposed a mix-precision approach where each layer of a DNN is composed of a low precision memristive array and a high precision digital part. The digital part computes and accumulates the weight update request in floating-point numbers, and the conductance of the elements in the memristive array is updated when the accumulated weight update request in the high precision digital part reaches a threshold. The memristive array performs the VMMs in an analog fashion and deals with the small input and output for error backpropagations, which requires high-performance DACs and ADCs. The high-precision digital part is more complex than our digital counter array since in our proposal the weight updates request (CD) only consists of integers. The comparison of the learning algorithm and training method with previously reported works of the memristive deep neural network can be seen in the Supplementary Information Table S1. According to the literature[14], the ADCs and DACs may account for 75% of the area and 87% energy consumption of the macro core consisting of the memristive array and peripheral circuits. Thus, a significant energy consumption reduction compared with the conventional design of memristive based deep neural network is expected.

Capacitor gated transistors[48] or other emerging electrolyte gate mem-transistors[50] with highly linear behaviors may also be used as the CD accumulation cells replacing the digital counters. Since the CD accumulation array is only needed in the training stage, it can be powered off at the inference stage and does not require long-term non-volatility.

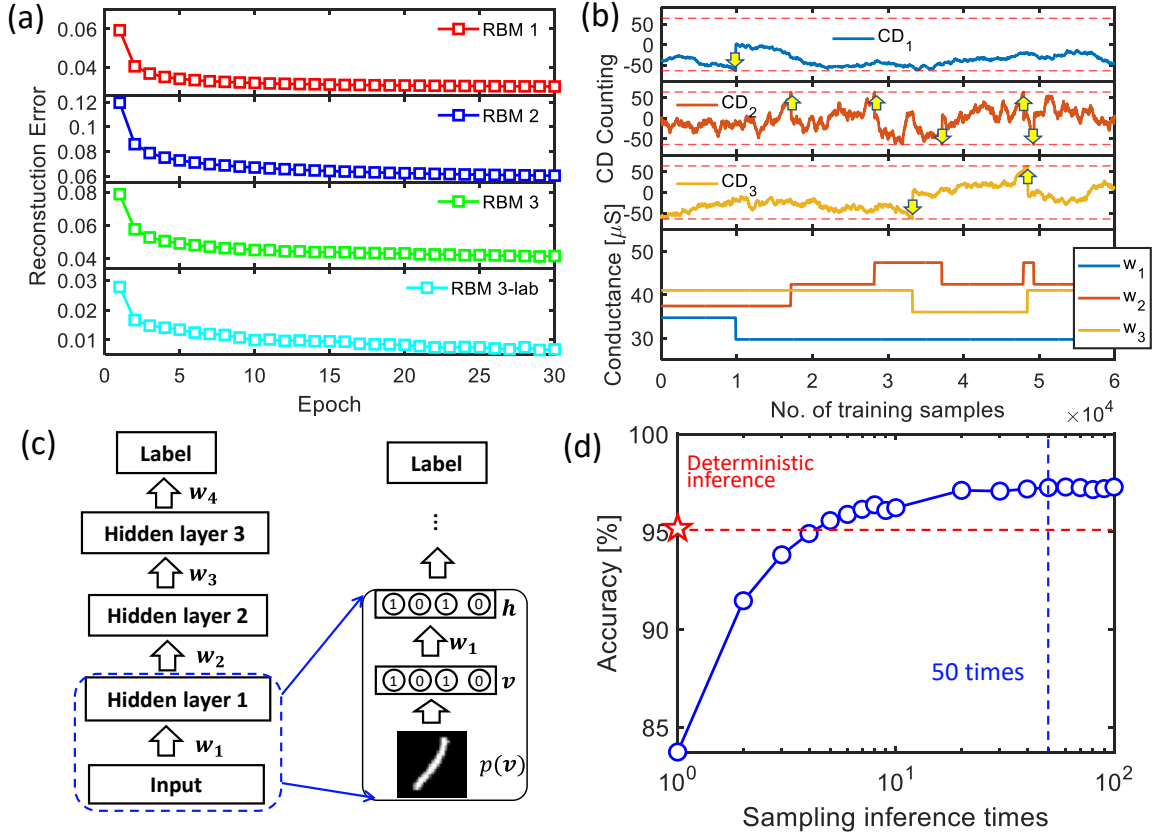

**Figure 2. Training and inference of the memristive DBN.** a) Reconstruction error for each RBM during training as a function of training epoch. b) Example evolution traces of digital CD counter and analog memristor as the function of the number of input training samples. c) Structure of the reorganized neural network for inference (pattern recognition). d) Comparison

of the accuracy between the fast deterministic inference and repeated sampling inference when using the well-trained DBN to recognize the handwritten digit images in the MNIST dataset.

## 3 Training and inference
### 3.1 Memristive DBN training

We first use a synaptic behavior with an ideally symmetric and linear weight update ability (Figure S1a) to test the applicability of the proposed training algorithm. The memristive DBN (Figure 1a) has the size of 784-500-(500+10)-2000, and each RBM is consecutively trained with the greedy learning algorithm for 30 epochs for all 60,000 images in the training set of the MNIST. **Figure 2a** shows the reconstruction error of visible neurons, defined as the normal distance between the original input and the reconstructed input $<|v'-v|>$, for each RBM during the training. The reconstruction error of the label neurons in the last RBM (RBM 3) is also shown in the figure. The gradually decreasing reconstruction error indicates that each RBM learned the input patterns to the visible neurons successfully. Three example evolutional traces of the value in the CD counter and the memristor conductance monitored during one training epoch (60,000 training samples) are shown in Figure 2b. From Figure 2b, we can see that the CD counting update when each training image is input, while most of them cancel each other and won't accumulate. When it reaches the threshold ($CD_{th}$=64) or is below the negative threshold (-64), the corresponding memristor conductance is potentiated or depressed, respectively. An animation is provided to show the greedy-learning process in a more illustrative way (Supplementary Movie 1).

Then, the DBN is fine-tuned by the wake-sleep algorithm for 30 epochs (Supplementary Movie 2, Figure S2),[51] which can be performed in the same hardware as in the greedy learning (see Methods for more details). Note that replacing the fully connected RBM layer with the convolutional RBM layer can effectively reduce the size of both the memristive array and the CD accumulation array resulting in better accuracy,[52] which, however, is beyond the scope of the current work. To scale-up the DBN for larger datasets, i.e., CIFAR-10, convolution RBM layers are also necessary.[53]

### 3.2 Inferences with binarized activations

The inference of the DBN, i.e., pattern recognition from the input to label, can be implemented by unfolding the last RBM, as shown in Figure 2c. Only forward VMM and stochastic sampling are needed. For a well-trained DBN, sequential implementation of the forward VMMs and stochastic sampling in each RBM layer results in reduced accuracy for all the test images in MNIST (~ 83.73%, Figure 2d). However, the accuracy can be gradually improved by repeating the sampling inference (~ 97.26% for 50 times repeats). The noise current can be disabled such that the excitation of each neuron is deterministic, i.e., deterministic inference. This results in an intermediate accuracy (~ 95.17%), however, at higher speed and lower power consumption. A trade-off between the slow-accurate and the fast-coarse inferences can be made according to application needs. Supplementary Figure S1b shows the test accuracies of deterministic inference and 50 times sampling inference (following training results will use this metric) as a function of training epochs.

### 3.3 Effect of the CD threshold

**Figure 3a** shows the performance of the training when varying $CD_{th}$ in the digital counter array (or varying the bit size of the signed counter). When $CD_{th}=1$ (corresponding to directly write to the memristor array for any immediate CD and no digital counter array), lower performance is observed. When $CD_{th}=64$ (6-bit counter), the highest recognition accuracy can be obtained. Further increase of the $CD_{th}$ will reduce the recognition accuracy since some weight requests are remained in the CD accumulation array and will not be transferred to the memristive array. Figure 3b shows the statistical results of the counts of write operations on each memristive cell for different $CD_{th}$. Without the digital counter array ($CD_{th}=1$), maximal $10^6$ and median $10^3$ write operations are performed on the memristive cells. When $CD_{th}=64$, the number of write operations reduced to maximal 200 and median 20, and half of the devices are not operated at all. The endurance specification for the memristive device is largely relaxed.

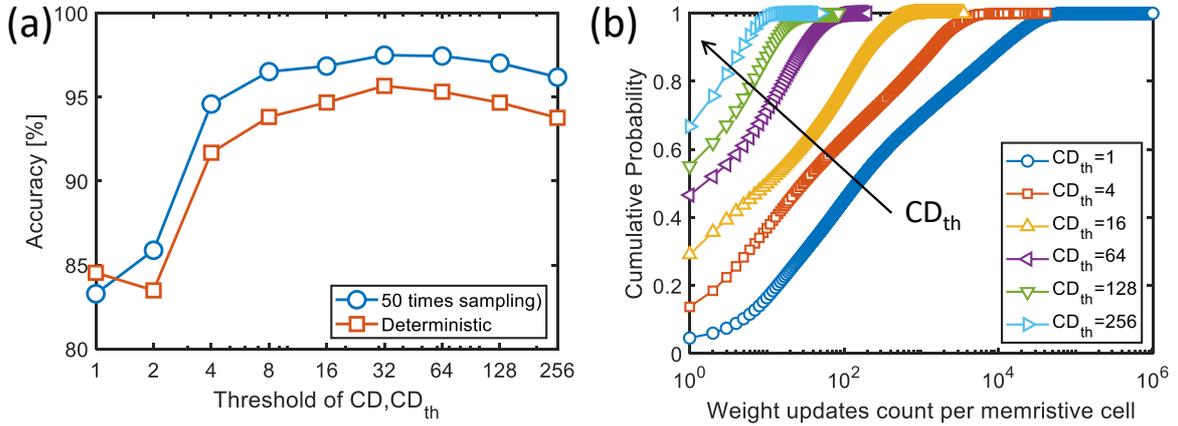

**Figure 3. Mixed-signal training of the DBN.** a) Test accuracy as a function of the training epoch for different $CD_{th}$ using the symmetric and linear weight update behavior. b) Statistical results of the counts of write operations on each memristive cell for different $CD_{th}$.

### 4. Immunity to non-idealities

To simulate more non-idealities of the memristive synaptic devices, an empirical model capturing conductance levels ($N_p$ and $N_d$ for potentiation and depression phases, respectively), on/off ratio ($G_{max}/G_{min}$), the non-linearities ($\alpha_p$ and $\alpha_d$), and the asymmetry between potentiation and depression ($N_p \neq N_d$, $\alpha_p \neq \alpha_d$) is proposed and shown in Figure 4a, which can be written as (without cycle-to-cycle and device-to-device variations),

$$\Delta G_{pot} = [\frac{G_{max}-G_{min}}{1-e^{-\alpha_p}} - (G - G_{min})](1 - e^{-\alpha_p/N_p}), \quad (4)$$

and,

$$\Delta G_{dep} = -[\frac{G_{max}-G_{min}}{1-e^{-\alpha_d}} - (G_{max} - G)](1 - e^{-\alpha_d/N_d}), \quad (5)$$

for potentiation and depression, respectively. Figure S3 shows the example traces of conductance evolution obtained from the model when random generated potentiation and depression pulses are applied. With this model in hand, we check the effects of various non-idealities of memristive devices on the performance of the memristive DBN.

### 4.1 Limited conductance levels

In contrast to the ideal analog conductance tunability, most memristive devices only show two conductance levels, i.e., low conductance state (LRS) and high conductance state (HRS).[2] Multiple conductance states are generally more promising in RRAM and PCM devices.[54] However, these multiple conductance states are usually obtained with external controlling stimuli, e.g., compliance currents or close-loop read-write-read verify technique.[22,55] Here, we simulate the case where the multiple conductance states are obtained by identical potentiation or depression pulse. The number of conductance levels is defined as the number of pulses ($N_p$) needed for the device to switch from the LRS to HRS in the potentiation phase or, vice versa, the number of pulses ($N_d$) needed for the device to switch from the HRS to LRS in the depression phase, as shown in Figure 4a.

Figure 4b shows the test accuracy after the training as a function of conductance levels. The network works well (accuracy >90%) even when only two conductance levels are available and reaches the best performance (accuracy >97%) when 20-40 conductance levels are available. The deterioration in the performance at higher conductance levels can be compensated if more training epochs are conducted (Figure 4b).

### 4.2 Non-linear weight update

Non-linear weight update behavior is another major source of performance lost when using memristive synaptic devices for the training of a neural network.[42,56] To verify the effect of weight update nonlinearity on the training of the memristive DBN, we vary the nonlinearities of both potentiation and depression in the model ($\alpha_p$ and $\alpha_d$) while keep them equal (Supplementary Figure S5a and S5b). Figure 4c shows that increasing the nonlinearity of the weight update will slightly decrease the training accuracy. Besides, increasing the $CD_{th}$ could partially compensate for the deterioration.

### 4.3 Asymmetric weight updates

The weight updates for potentiation and depression generally do not have the same degree of non-linearity, i.e., asymmetric nonlinear weight updates. To test the effect of asymmetric weight updates, we fix the non-linearity for the depression phase ($\alpha_d$) and only vary the non-linearity for the potentiation phase ($\alpha_p$) (Supplementary Figure S5c and S5d). Figure 4d shows the performance of the memristive DBN as a function of the asymmetry between the weight update non-linearities of potentiation and depression phases.

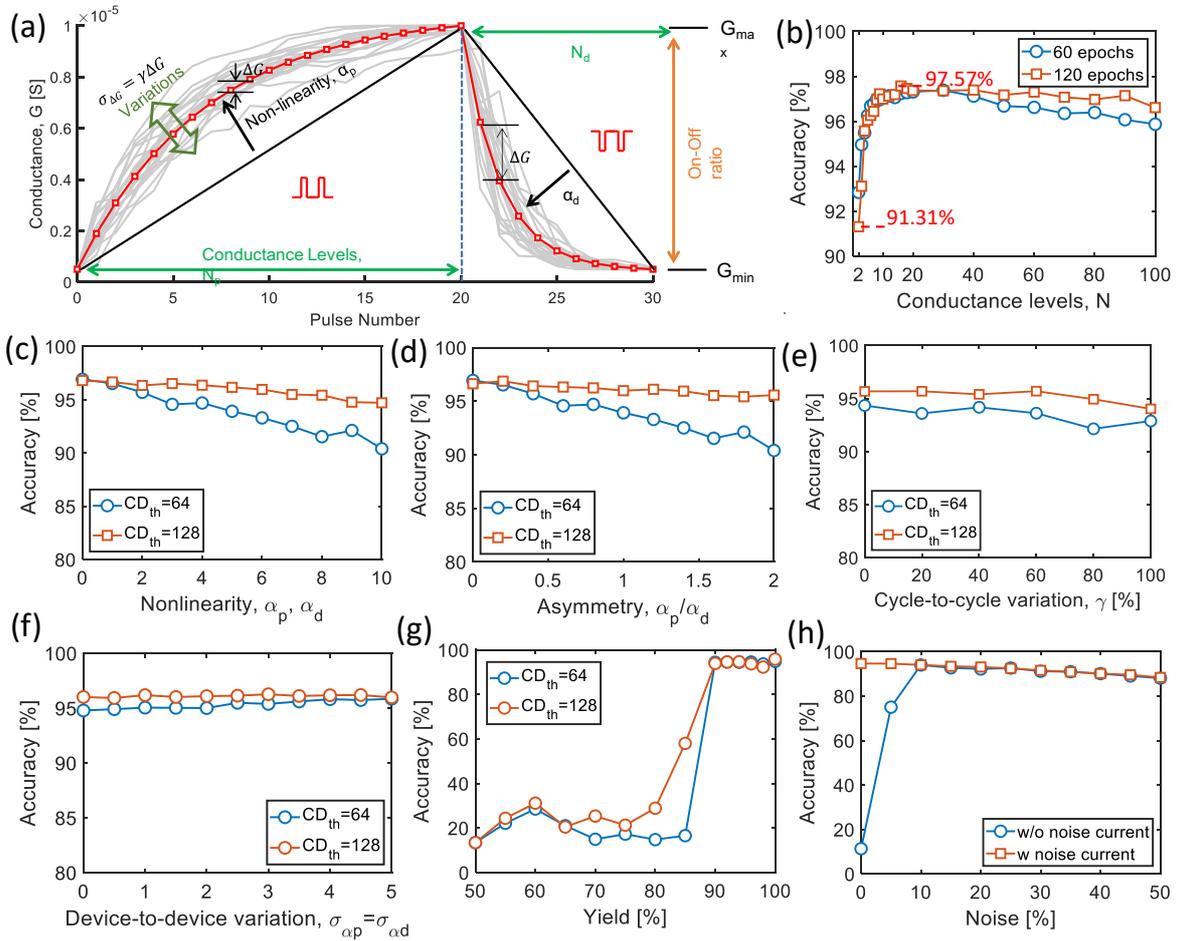

**Figure 4. Effect of non-idealities of the synaptic device on the training performance of the memristive.** a) An empirical model to capture more non-idealities of the memristive synaptic devices: nonlinear weight update, the asymmetry between potentiation and depression, and write variation. (Red lines: model w/o variations; Gray lines: model with variations). Training accuracies as b) a function of conductance levels, c) the symmetric nonlinearity, d) asymmetric

nonlinearity, e) cycle-to-cycle variation, f) device-to-device variation, g) yield, and h) read noise.

**4.4 Write variations**

Another source of performance degradation of the memristive neural network comes from the cycle-to-cycle and device-to-device variations when write to memristive devices.[57,58] The cycle-to-cycle write variations are modeled by adding a Gaussian distribution to the conductance change with its standard deviation proportional to the ideal conductance change for each weight update operation,

$$\Delta G_{c2} \in \mathcal{N}(\Delta G, \sigma_{c2c}^2), \ \sigma_{c2c} = \gamma \Delta G. \tag{6}$$

(Supplementary Figure S6a and S6b). The simulation result shows that in the proposed memristive DBN, the cycle-to-cycle write variations only slightly affect the test accuracy of the neural network (Figure 4e).

Device-to-device variation is modeled by assigning each device's weight update non-linearity according to a Gaussian distribution [$\alpha_{p,d2d} \in \mathcal{N}(\alpha_p, \sigma_{\alpha p}^2)$ and $\alpha_{n,d2d} \in \mathcal{N}(\alpha_n, \sigma_{\alpha n}^2)$, Supplementary Figure S6c and S6d]. Figure 4f shows the recognition accuracy of the memristive DBN as a function of the standard deviation of the device's non-linearity. Surprisingly, the higher device-to-device variation does not degrade the performance of the neural network. We see a slight increase in the recognition accuracy.

**4.5 Device yield**

In memristive devices, especially the RRAM devices, device yield is the major issue preventing its application in data storage and neuromorphic computation on a large scale.[55,59] The memristive device may not work due to the process variation or in some other cases the synaptic devices may initially work well but then stuck at HRS or LRS during the following write operations. In the simulation shown in Figure 4g, we assume that a percentage of the devices are not working (half of them stuck in HRS and the other half stuck in LRS). From Figure 4g, we see that when the device yield is higher than 90%, the performance of the memristive DBN does not degrade. While when the yield is less than 90%, the accuracy of memristive DBN training quacking drops to 20%. Two factors cause the accuracy drop for low device yield: 1) low device yield prevents the accurate greedy learning layer-by-layer; 2) the fine-tuning after the greedy learning is more sensitive to the nonidealities of the memristive devices thus ruin the previously learned recognition ability through greedy learning algorithm (Supplementary Information Figure S7a).

**4.6 Read noise**

Multiple sources of noise can induce inaccuracy in the reading of the memristive devices, for instance, flicker noise, random telegraphy noise, and white noise, etc.[60–62] The noise read instability could also be originated from the sense amplifiers and other peripheral circuits. The inaccurate read current will result in the inaccurate output of the VMM. However, since the proposed memristive DBN has stochastic output, the read noise could be a beneficial factor making the hardware implementation easier. As discussed earlier, the probabilistic behavior of the neurons in RBM induced by the noise current injected to the input of each column of the memristive array in Figure 1c can be realized by properly tuning the signal to noise ratio of the memristive device reading.[47] Here, we test the effect of reading noise by adding a current noise in each of the memristive devices and test two cases, i.e., without and with noise current injected to the neuron. Figure 4i shows the performance of the memristive DBN as a function of the read noise level for the two cases. With noise current injected into the neuron, i.e., probabilistic neurons as designed earlier, the read noise of the memristive device slightly lowers the training performance. While, without noise current injected into the neuron, when the read noise is small, the memristive DBN shows highly degraded recognition accuracy after training. A certain noise level will be beneficial to the neural network.

**4.7 Nonlinear I-V characteristic**

Another common non-ideal behavior of the memristive device is its non-linear I-V characteristics.[13,63] This prevents the direct implementation of the multiplication in the analog domain since the conductance of the device is not a constant value at different read voltages. Pulse width or pulse number modulation is usually used to represent the analog input of VMM in the implementation of a neural network.[18,64] Method of operating the devices in a small dynamic range to avoid the non-Ohmic conduction or in the small signal domain has also been proposed.[65,66] All the solutions come with the price of complexing the readout circuit for VMM operations. In the proposed memristive DBN structure, however, the input of the VMM is also binary-valued, which is inherently immune to the non-linear I-V characteristic issue of memristive devices.

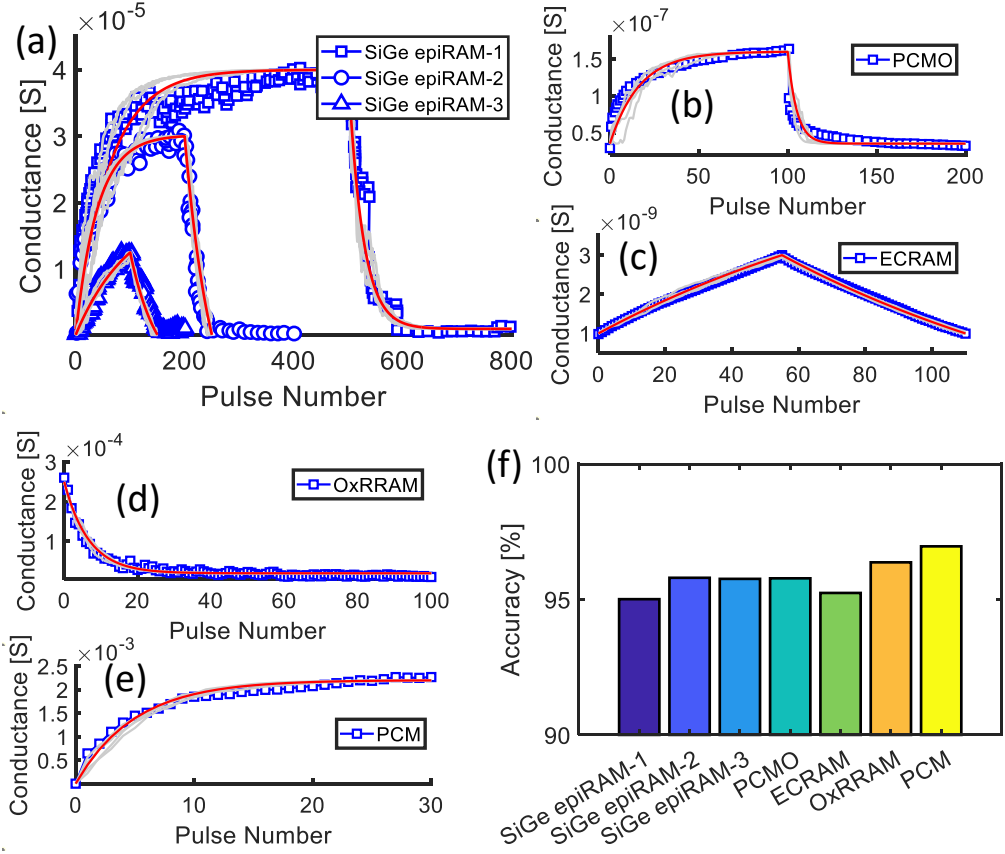

**Figure 5. Training of the memristive DBN with memristive synaptic devices data**. Fitting memristive synaptic behaviors by the device model for a) SiGe epiRAM, b) PCMO, c) ECRAM, d) OxRRAM, and e) PCM, respectively. Blue points: data from the references; red lines: model w/o variations; gray lines: model with variations. f) Training accuracy of the memristive DBN using the memristive device behaviors ($CD_{th}$=128).

## 5. Memristive DBN with real memristive synaptic devices

The synaptic model in Figure 4a is used to fit various memristive synaptic behaviors of SiGe epiRAM[67], PCMO[23], ECRAM[50], OxRRAM[68], and PCM[69] devices (**Figure 5a-e** and Supplementary Table S2). The fitting parameters are used to validate the training of the memristive DBN. Since device-to-device variations and device yield data are not shown in the references, thus we assumed ideal parameters for them. For the devices that only have gradual weight update behavior in one direction, i.e., OxRRAM[68], and PCM[69], we use the differential pair where each synaptic cell contains two devices (Supplementary Figure S8). All the memristive synaptic behaviors obtained by identical potentiation and depression pulses enable the successful training of the memristive DBN with accuracies ranging from 95% to 97% (Figure 5f and Supplementary Figure S9). Note that in Section 4, to validate the learning algorithm accommodate to all synaptic behavior, the simulation includes the extreme cases where the nonidealities of the memristive devices are unusually high. In these cases, the fine-tuning would ruin the previously learning recognition ability (Figure S4, S5, and S7). For the synaptic behavior of real memristive devices, as shown in Figure 5 and Figure S9, we see that the fine-tuning always improves the performance of the neural network.

## 6. Relaxed specifications for memristive synapses

By properly balancing among the parameters of various nonidealities, we got a set of parameters required for the memristive synaptic devices that can achieve 95% accuracy for the MNIST dataset, as listed in Table 1. The specifications of memristive devices in the literature [26,42–44] are also listed in the table to make the comparison. The proposed memristive DBN has high relaxed specifications for memristive devices. According to our simulation experience of checking various nonidealities, we found that the non-linearity of weight update is the most important factor need to be taken care of. Reducing the non-linearity would continuously improve the performance of the neural network. The number of conductance levels and device yield need special care since a low number of conductance levels and low device yield will suddenly deteriorate the performance of the neural network. However, when enough conductance levels and device yield are available, further improvement of these metrics would not significantly benefit the neural network's performance. Other nonidealities are highly relaxed and can be easily meet with most of the existing memristive device technology.

**Table 1. Specifications of memristive devices to achieve 95% accuracy for training MNIST compared with the specifications in the literature.**

|  | T. Gokmen, et al., 2016[26] | P.Y. Chen, et al., 2017[43] | C.C. Chang, et al., 2018[42] | S. Agarwal, et al., 2016[44] | This work |
|---|---|---|---|---|---|
| Conductance levels | ≥1000 | ≥64 | ≥256 (8 bits) | - | ≥20 |
| On/off ratio | ≥8 | ≥50 | - | - | ≥3 |
| Nonlinearity | - | ≤1.0 | ≤4.5[b] | ≤5 | ≤10 |
| Asymmetry | ≤1.05 | ≤2 % | - | - | ≤2[a] |
| Cycle-to-cycle variation | - | - | - | ≤0.4 % | ≤30 % |
| Device-to-device variation | - | ≤1 | - | - | ≤5 |
| Yield | - | - | - | - | ≥90 % |
| Read noise | - | ≤20 % | - | ≤9 % | ≤10 % |
| Endurance | - | - | - | - | ≥100 |
| DAC/ADC accuracy | 9 bits | - | 8 bits | - | 1 bit |
| Nonlinear activation | Software | Software | Software | Digital core | Not required |

'-': Not discussed; [a] Could be removed if differential pairs are used; [b] Converted from the original value since different metric of non-linearity is used in the reference.

## 7. Conclusion

A memristive DBN composed of mixed-signal RBM layers for efficient online training is proposed. The mixed-signal RBM layer consists of an analog memristive array for the stochastic

VMM and a digital counter array for the accumulation of CD. The proposed memristive DBN has stochastically binarized activation, free from the need for complex peripheral circuits with expensive DACs and ADCs. It shows high immunity to various non-idealities of the memristive synaptic devices. The endurance requirement of the memristive is also highly relaxed.

## 8. Methods

*Training of a single RBM layer*: We used the first order CD for the training of memristive RBM. The input (state of the visible units, $\boldsymbol{v}$) is first multiplied by the weight matrix of the RBM 1 ($\boldsymbol{w_1}$) to obtain the probability of the state of the hidden units ($\boldsymbol{h}$),

$$P(\boldsymbol{h} = 1) = \sigma(\boldsymbol{v}\,\boldsymbol{w_1}), \tag{7}$$

where $\sigma(x) = \frac{1}{1+e^{-x}}$ is the logistic sigmoid function. Then the state of the hidden units ($\boldsymbol{h}$) is backward multiplied by the weight matrix to obtain the probability of the reconstructed state of the visible units ($\boldsymbol{v'}$),

$$P(\boldsymbol{v'} = 1) = \sigma(\boldsymbol{h}\,\boldsymbol{w_1^T}). \tag{8}$$

After that, the reconstructed state of the visible units ($\boldsymbol{v'}$) is again multiplied by the weight matrix to obtain the probability of the reconstructed state of the hidden units ($\boldsymbol{v'}$),

$$P(\boldsymbol{h'} = 1) = \sigma(\boldsymbol{v'}\,\boldsymbol{w_1}). \tag{9}$$

The CD is then calculated by the difference between the outer-products of the two sets of visible-hidden neuron states,

$$\boldsymbol{CD} = \boldsymbol{v} \otimes \boldsymbol{h} - \boldsymbol{v'} \otimes \boldsymbol{h'} \tag{10}$$

The CD matrix which acts as the gradient descent of the weight matrix is used to update the weight matrix,

$$\frac{\partial E}{\partial w_1} \propto \boldsymbol{CD}, \tag{11}$$

where $E$ is the energy of the RBM which should be minimized. In this work, we do not update the weight matrix $\boldsymbol{w_1}$ directly. We accumulate the CD matrix (request for weight updates), and only update the elements in the weight matrix when the elements in the CD accumulation (signed integers stored in digital counters) reach a threshold.

*Training of the last RBM with label input*: When training the last RBM, label target ($\boldsymbol{l}$) in one-hot format is input to the label layer with 10 neurons. The weight matrix is partitioned into two parts ($\boldsymbol{w_3}$ and $\boldsymbol{w_4}$ in Figure 1a). The forward VMM and sampling are performed by both inputting the hidden neuron states of the 2nd RBM to the visible layer and inputting the target label into the label layer. The hidden neurons sum all the currents both from the visible layer and the label layer. The backward VMM and sampling are performed separately from the hidden layer to the visible layer via weight matrix $w_3$, and from the hidden layer to the label layer via weight matrix $w_4$. The neurons in the visible layer are excited according to the probability given by the sigmoid function as described in Eq. 7. The neurons in the label layer, however, are excited according to the probability given by the softmax function,

$$P(l' = 1) = SoftMax(h\,w_1^T), \qquad (12)$$

where $SoftMax(x)_i = \frac{e^{-x_i}}{\sum e^{-x_i}}$, which makes sure that, statistically, only one label neuron will be excited. The CD matrix and its accumulation will be implemented separately for the weight matrices $w_3$ and $w_4$, respectively.

*Week-sleep algorithm for fine-tuning*: The fine-tuning of the memristive DBN by the week-sleep algorithm is performed after the greedy-learning of the DBN (pretraining). To do the fine-tuning, the RBMs in the memristive DBN are duplicated except for the last RBM with new weight matrices $w_1 = w_1'$ and $w_2 = w_2'$ (Figure S2a). The weight matrices $w_1$, $w_2$, $w_3$, and $w_4$ constitute the recognition path, and the weight matrix $w_4$, $w_3$, $w_2'$, and $w_4'$ constitute the generation path or wake path. The input data of the image and the label are feed to the recognition path, and the top RBM layer perform Gibb sampling iteratively (iteration number is 20 in our simulation), and then the reconstructed visible neuron states are feed to the generation path to generate the reconstructed image (sleep path). The states of neurons in the wake path and sleep path are used to update the weight matrix in the sleep path and wake path, respectively (Figure S2b). In this paper, the DBN is pre-trained for 30 epochs, and fine-tuned for another 30 epochs. From the detailed recognition accuracy vs epoch traces in Supplementary Figure S4-S7, we could see that the pretraining always improves the performance of the neural networks, while fine-tune can further improve the performance in some cases. When the non-idealities of the memristive device are high, the fine-tuning is not working well or may ruin the previously learned recognition ability. This indicates that the fine-tuning procedure is more prone to device non-idealities. The highest accuracy in the traces of the accuracy versus the epoch is taken as the metric for the analysis in Figure 3, Figure 4, and Figure 5.

*Image generation by the trained DBN:* The memristive DBN after fine-tune with generation path can be used to generate image when only given label input (Figure S10). As shown in Figure S10a, a random noise image and a label are input to the DBN, and the top RBM layer performs the Gibbs sampling in multiple iterations. Taking the reconstructed visible neuron states of the top RBM layer as input, the generation path's output will provide the correct digits image corresponding to the label (Figure S10b).

**Supporting Information**

Supporting Information is available from the Wiley Online Library or from the author.

**Acknowledgements**

This work was supported by the European Research Council through the European Union's Horizon 2020 Research and Innovation Programe under Grant 757259. This project has also received funding from theEuropean Union's Horizon 2020 Research and Innovation ProgrammeFET Open NEU-Chip under grant agreement No. 964877. W.W. was supported in part at the Technion by the Aly Kaufman Fellowship. The Interactive Supporting Information of this article can be found at: https://www.authorea.com/doi/full/10.22541/au.164130702.22387391.

ToC

An efficient online training of memristive deep belief net (DBN) with CD algorithm is proposed. The memristive DBN needs no DACs or ADCs and shows high immunity to various non-idealities of synaptic devices. 95%~97% recognition accuracies are achieved for MNIST by various synaptic devices. Specifications for synaptic devices are highly relaxed.

W. Wang, B. Hoffer, T. Greenberg-Toledo, Y. Li, M. Zou, E. Herbelin, R. Ronen, X. Xu, Y. Zhao, J. Yang, and S. Kvatinsky*

Efficient Training of the Memristive Deep Belief Net Immune to Non-Idealities of the Synaptic Devices

ToC figure

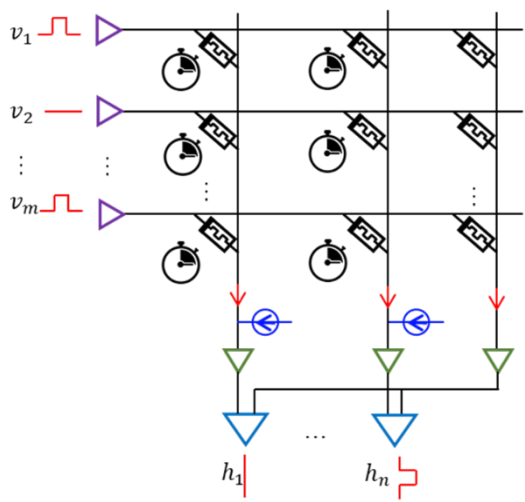

# Supporting Information

Efficient Training of the Memristive Deep Belief Net Immune to Non-Idealities of the Synaptic Devices

W. Wang, B. Hoffer, T. Greenberg-Toledo, Y. Li, M. Zou, E. Herbelin, R. Ronen, X. Xu, Y. Zhao, J. Yang, and S. Kvatinsky*

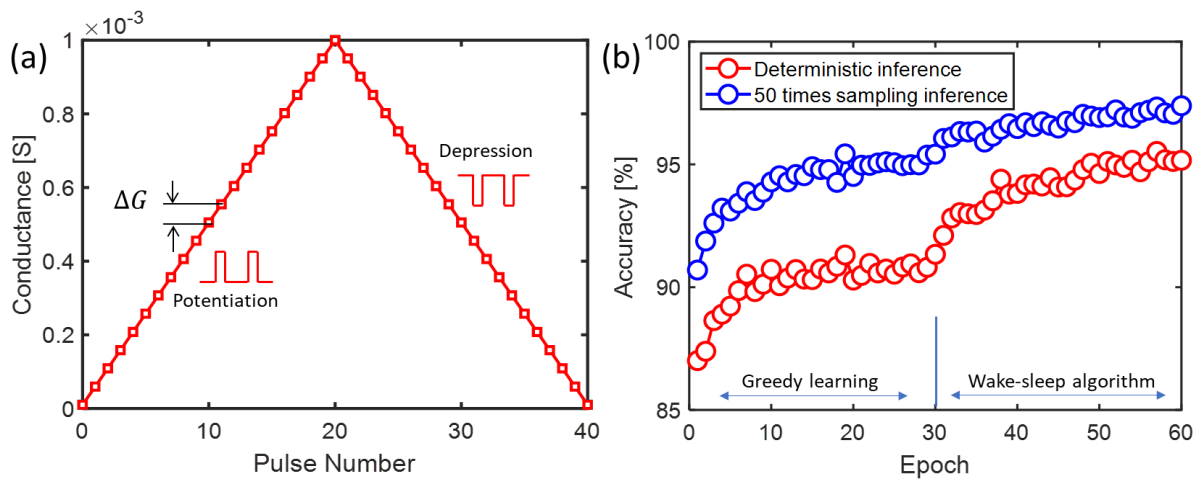

**Figure S1.** a) The memristive synaptic behavior with an ideally symmetric and linear weight update ability (constant ΔG for identical pulses) but limited conductance levels ($N$=20). b) Test accuracy for 10,000 images in the MNIST dataset obtained during the training of memristive DBN as a function of the training epoch ($CD_{th}$=64).

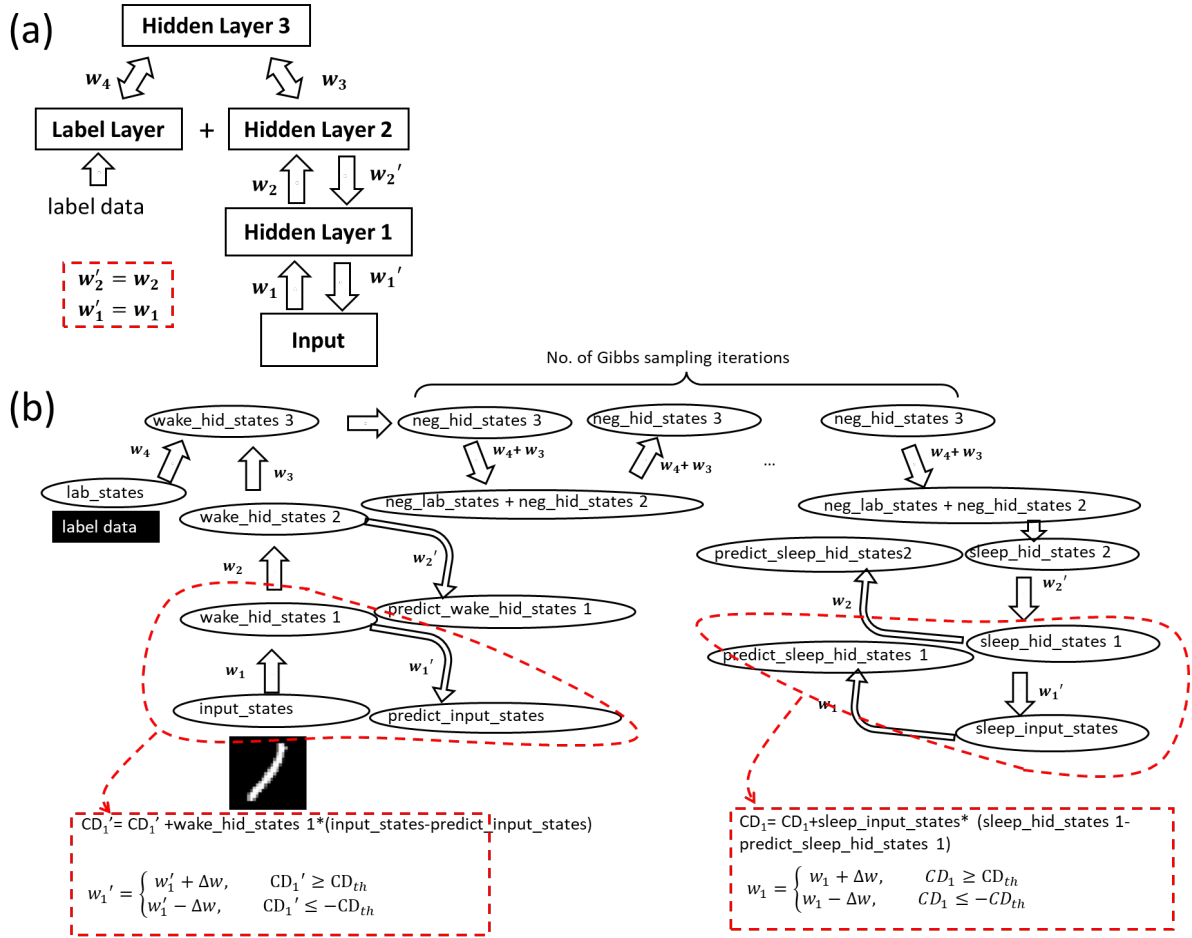

**Figure S2.** a) Deep belief network structure for the fine-tuning procedure by the wake-sleep algorithm. b) Procedure of utilizing wake-sleep algorithm for the fine tune of the deep belief net.

**Table S1.** Comparison of the training method with previous works

|  | T. Gokmen, et al., 2016[26] | S. Ambrogio, et al., 2018[48] | P. Yao, et al., 2020[14] | This work |
| --- | --- | --- | --- | --- |
| Memory cell | 1R | 2PCM+3T1C | 1T1R | 1R+Digtial Counter |
| Training algorithms | Error Backpropagation | Error Backpropagation + Weight transfer | Weight transfer + in-situ training of the last layer | Contrastive Divergence |
| Weight update method | Stochastic blind write | Blind update for 3T1C; closed-loop update for 2PCM | Closed loop | Blind write |
| Tested dataset | MNIST | MNIST, CIFAR-10, CIFAR-100 | MNIST | MNIST |
| Recognition accuracy (for MNIST) | 99% (for ideal memristive device) | 97.95% | 96.92% | 97% |
| DAC/ADC accuracy | 9 bits | - | 8 bits | 1 bit |
| Nonlinear activation | Software | Software | Software | Not required |

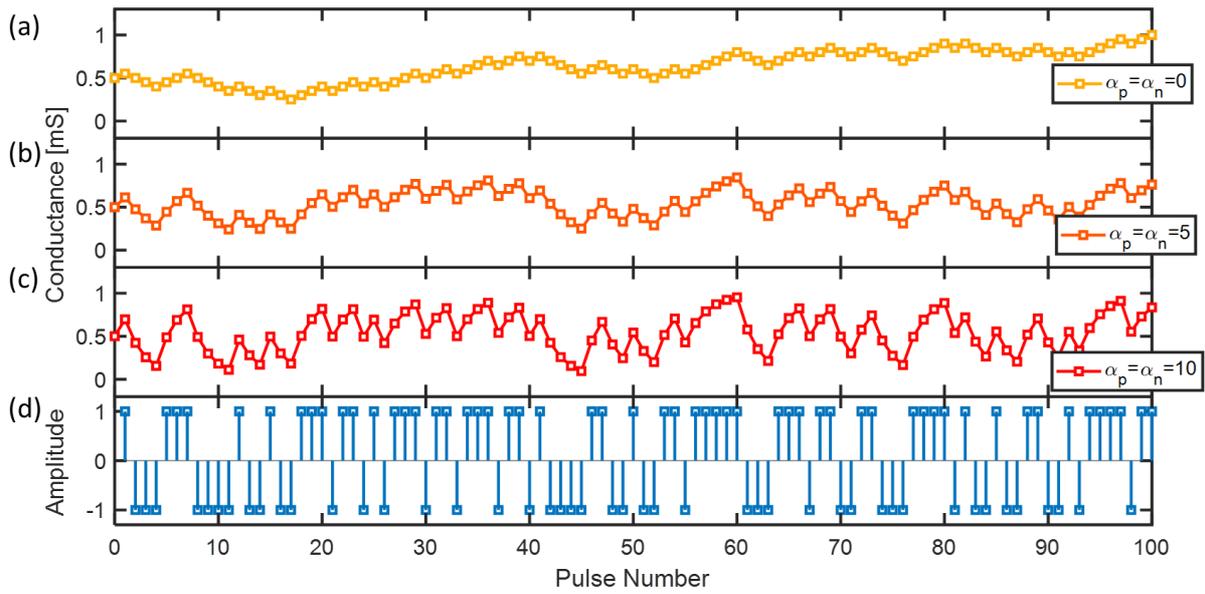

**Figure S3.** Example traces of the conductance updates as the responses of potentiation or depression voltage pulses. a) Weight update traces of an ideal device with linear weight update ability. b, c) Weight update responds of an ideal device with non-linear weight update behavior. d) Potentiation (positive) or depression (negative) pulse applied on the devices.

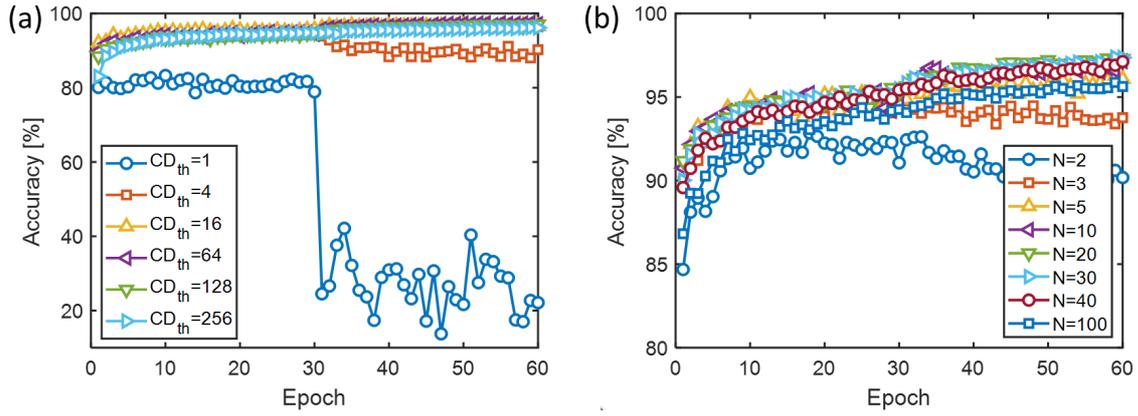

**Figure S4.** a) Test accuracy (50 times sampling) as a function of the training epoch for different CD$_{th}$ using the symmetric and linear weight update behavior. b) Test accuracy as a function of the training epoch for different numbers of conductance levels, N=N$_p$=N$_d$.

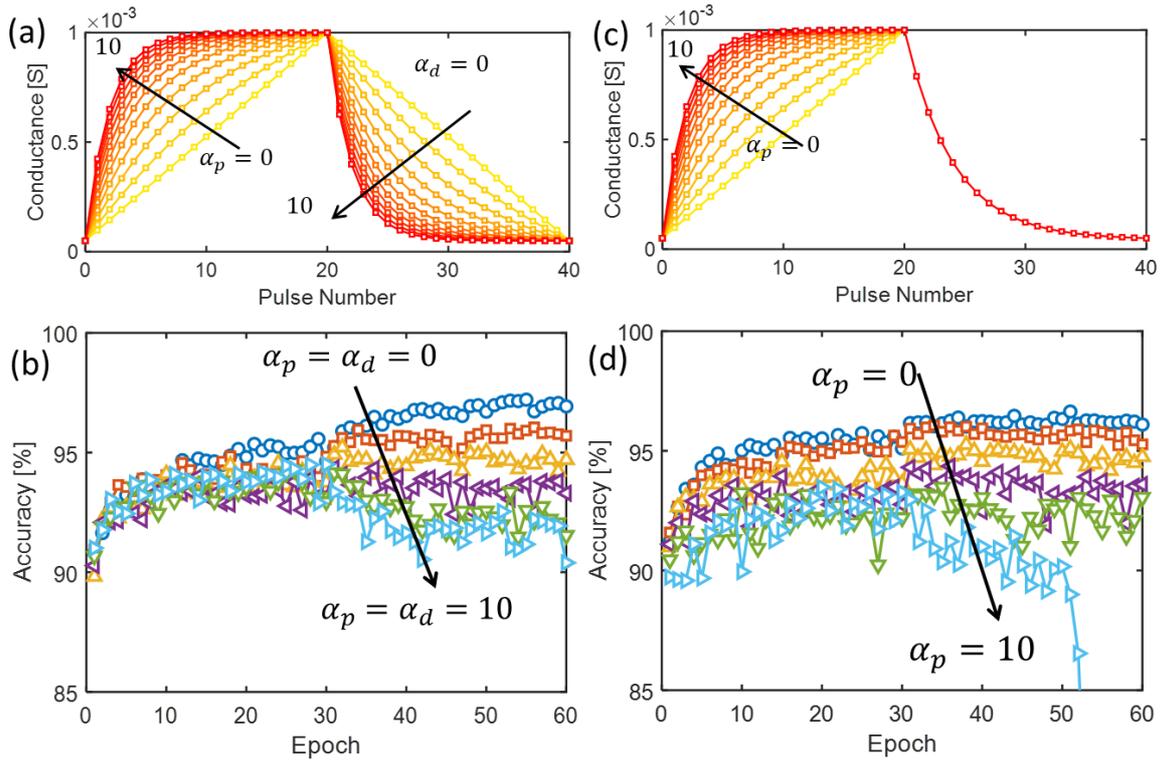

**Figure S5.** a) Modelled weight update behavior with different degrees of non-linearities (assuming the degrees of non-linearity in the potentiation and depression are the same). b) Test accuracy as a function of the training epoch for different degrees of symmetric nonlinearities of the weigh updates. c) Modelled weight update behavior with different degrees of non-linearities (assuming the degree of non-linearity in depression is constant, i.e., asymmetric weigh updates). d) Test accuracy as a function of the training epoch for different degrees of asymmetric nonlinearities of the weigh updates.

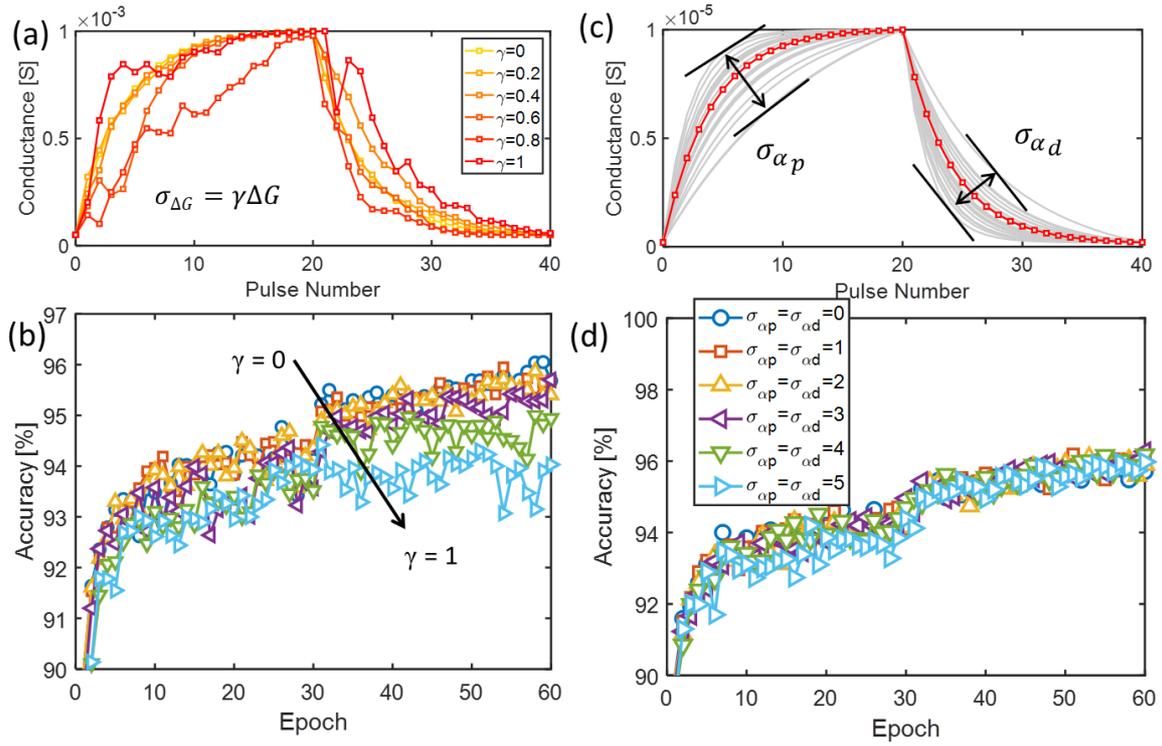

**Figure S6.** a) Modelled weight update behavior with different degrees of cycle-to-cycle variations. b) Test accuracy as a function of the training epoch for different degrees of cycle-to-cycle variations. c) Modelled weight update behavior with different degrees of device-to-device variations. d) Test accuracy as a function of the training epoch for different degrees of device-to-device variations.

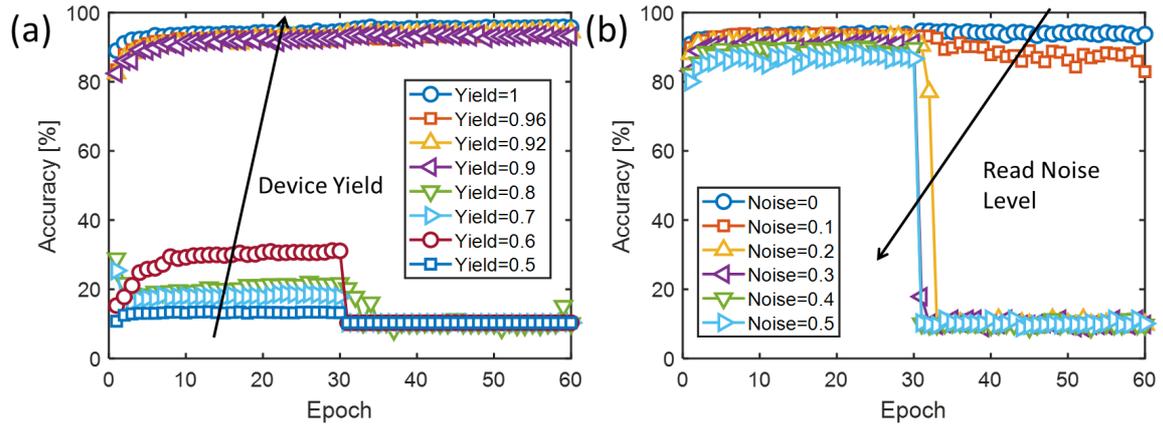

**Figure S7.** a) Test accuracy as a function of the training epoch for various yields of the memristive device in the array. b) Test accuracy as a function of the training epoch for various read noise levels of the memristive device.

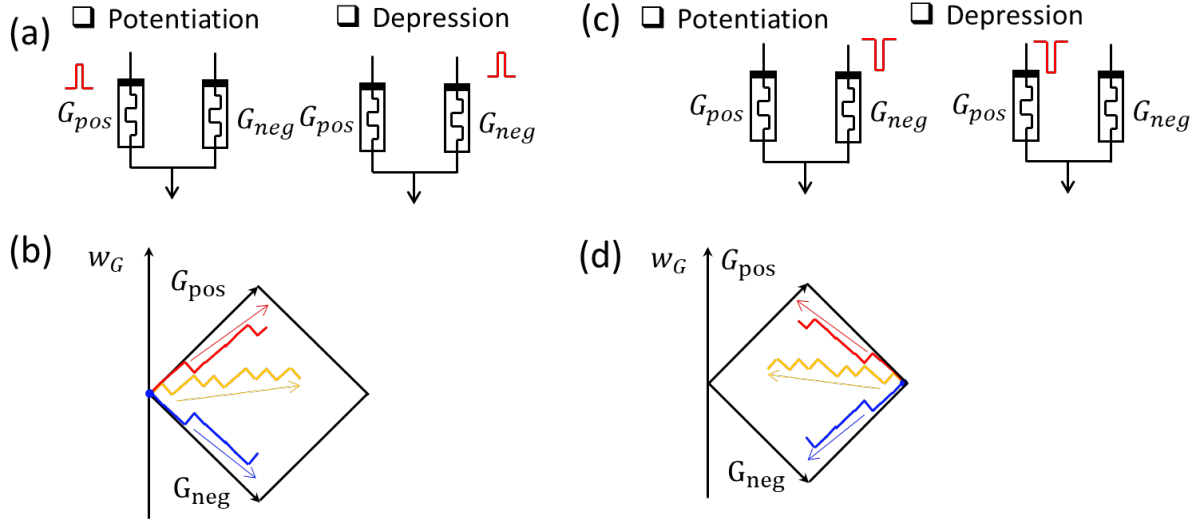

**Figure S8.** a) For the device that only has gradual conductance incremental behavior (PCM device in Fig. 5e), potentiation and depression operation are performed in the positive branch and negative branch of the synaptic cell consisting of a differential pair of the memristive devices, respectively. b) Illustration of typical synaptic evolution of the differential pair. The two memristive devices are initialized to low conductance status. Increment of the conductance of the positive branch results in the gradual increase of the overall synaptic weight. Vice-versa increment the conductance of the negative branch results in the gradual decrease of the overall synaptic weight. However, after multiple weight update operations, the differential pair might reach a status that both the devices are in high conductance states, which prevents further weight update operations and degrades the performance of the neural network. In the training case in this work, no such issues were detected, since the number of weight update operations on each memristive synaptic cell is highly reduced. c) For the device only has gradual conductance decrement behavior (OxRRAM device in Fig. 5d), potentiation and depression operation are performed in the negative branch and positive branch of the synaptic cell consisting of a differential pair of the memristive devices, respectively. b) Illustration of typical synaptic evolution of the differential pair.

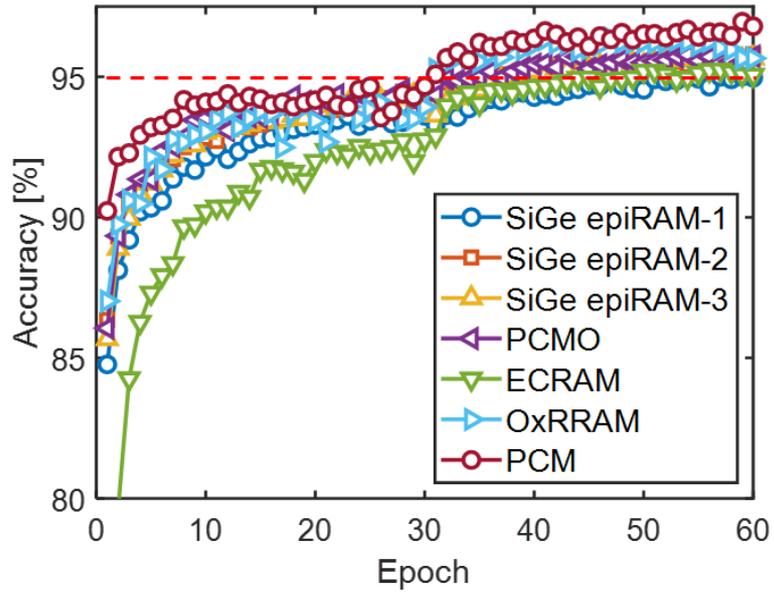

**Figure S9.** Test accuracy as a function of the training epoch when using various device data to train the neural network.

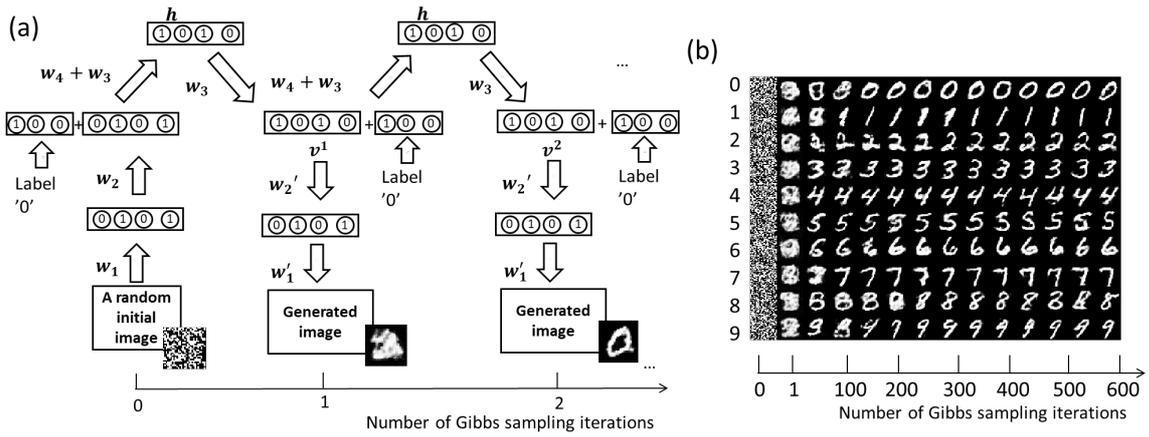

**Figure S10.** a) Schematic of using the trained DBN to generate handwritten digit images by only providing the labels. b) Generated handwritten digit images by providing only the labels.

**Table S2.** The fitting parameter values for the various memristive synaptic behaviors.

| Device | $G_{max}$ | $G_{min}$ | $N_P$ | $N_d$ | $α_p$ | $α_d$ | $γ$ | Accuracy |
|---|---|---|---|---|---|---|---|---|
| SiGe epiRAM [9] | 40 µS | 1 µS | 500 | 400 | 8 | 15 | 2 | 95.01% |
| | 30 µS | 0.1 µS | 200 | 50 | 5 | 1 | 1 | 95.80% |
| | 12.5 µS | 0.1 µS | 100 | 50 | 1 | 1 | 1 | 95.76% |
| PCMO [10] | 0.16 µS | 35 nS | 100 | 100 | 6 | 20 | 1 | 95.78% |
| ECRAM [6] | 3 nS | 1 nS | 55 | 55 | 0.5 | 0.5 | 0.3 | 95.24% |
| OxRRAM [11] | 250 µS | 20 µS | - | 100 | - | 15 | 0.3 | 96.37% |
| PCM [12] | 2.2 mS | 7 µS | 30 | - | 6 | - | 0.3 | 96.96% |